\documentclass[twocolumn,prl]{revtex4}
\usepackage{amssymb}

\usepackage{graphics}
\usepackage{epsfig}

\usepackage{dcolumn}
\usepackage{amsmath}
\usepackage{color}
\usepackage{ulem}

\begin{document}

\title{Competing `soft' dielectric phases and detailed balance in thin film manganites}
\author{Patrick R. Mickel, Amlan Biswas, and Arthur F. Hebard}
\affiliation{Department of Physics, University of Florida, Gainesville, FL 32611, USA}
\date{\today}
\begin{abstract}

Using frequency dependent complex capacitance measurements on thin films of the mixed-valence manganite (La$_{1-y}$Pr$_{y}$)$_{1-x}$Ca$_{x}$MnO$_{3}$, we identify and resolve the individual dielectric responses of two competing dielectric phases. We characterize their competition over a large temperature range, revealing they are in dynamic competition both spatially and temporally. The phase competition is shown to be governed by the thermodynamic constraints imposed by detailed balance. The consequences of the detailed balance model strongly support the notion of an `electronically soft' material in which continuous conversions between dielectric phases with comparable free energies occur on time scales that are long compared with electron-phonon scattering times.
\end{abstract}
\maketitle

\section{I. Introduction}
Phase separation and phase competition are associated with many of the most exotic material properties that complex oxides have to offer and are found ubiquitously in high-temperature superconductors\cite{HTSC,PStJ}, spinels\cite{Spinel}, multiferroics\cite{MultiF,MultiF2}, and mixed-valence manganites\cite{MPS,MFM}.  Accordingly, understanding the fundamental mechanisms of phase separation/competition is necessary for the technological implementation of these next generation materials.  In mixed-valence manganites, the disorder\cite{Disorder} and strain\cite{Strain} based explanations have recently been augmented by a model describing an \lq\lq electronically soft\rq\rq coexistence, where the phase separation is driven by delocalized thermodynamic physics\cite{ESP}. This theory has broad implications for complex oxides with coexisting and competing phases\cite{dagotto,banerjee,nucara}, however, evidence for `electronically soft' phases has yet to be provided. 

In this report we utilize frequency-dependent dielectric measurements of thin films of the mixed phase manganite, (La$_{1-y}$Pr$_{y}$)$_{0.67}$Ca$_{0.33}$MnO$_{3}$ (LPCMO), to separately identify charge ordered insulating (COI) and paramagnetic insulating (PMI) phases and then to provide a spatial and temporal description of the dynamic competition between these phases over a broad temperature range. We find that the constraints imposed by detailed balance strongly support the notion of an `electronically soft' material, as we observe continuous conversions of dielectric phases with comparable free energies competing on time scales that are long compared with electron-phonon scattering times.

\section{II. Experimental}

\subsection{A. Sample Fabrication}

Our sample geometry comprises a (110) oriented NdGaO$_{3}$ substrate, a 30 nm thick epitaxial thin film (La$_{1-y}$Pr$_{y}$)$_{0.67}$Ca$_{0.33}$MnO$_{3}$ (with $y \approx 0.5$) bottom electrode grown by pulsed laser deposition, a 10 nm thick AlO$_{x}$ dielectric grown by RF-sputtering, and a 50 nm thick Al top electrode grown by thermal deposition (see the inset of Fig. 1a). The (110) oriented substrate was chosen since it is well lattice matched to LPCMO. Four additional samples with thicknesses in the range 30~nm to 150~nm have shown similar results to those reported here. For further details on fabrication of the (La$_{1-y}$Pr$_{y}$)$_{0.67}$Ca$_{0.33}$MnO$_{3}$ films see Ref.~\cite{growth}.

\subsection{B. Impedance Measurements}

The \textit{a}-\textit{b} plane resistance of the LPCMO film was measured using a four probe geometry, sourcing current and measuring voltage. The temperature dependence of the resistance is shown in Fig.~1, demonstrating that with decreasing temperature $T$ the resistance increases smoothly until $T = 115$~K, and then decreases as an expanding ferromagnetic metallic (FMM) phase forms a percolating conducting network\cite{MPS,MFM} at the expense of the insulating dielectric phases. In bulk LPCMO samples there is also a signature kink in $R(T)$ in the range 200-220~K (interpreted as the temperature where the COI phase becomes well established\cite{TCO1,TCO2,COI_T} which is absent here, suggesting the COI phase is not present. However, as described below, our complex capacitance measurements demonstrate an increased sensitivity to dielectric phases compared to the dc resistance, and convincingly confirm the presence of the COI phase. 

Dielectric measurements are made on the LPCMO film using the trilayer configuration discussed above in which the manganite serves as the base electrode. Using this technique, which enables the study of leaky dielectrics by blocking shorting paths (see Ref.~\cite{CMC} and Sec.~\ref{sec:circuit}III\thinspace A for a detailed discussion), we measure the complex capacitance over the bandwidth 20~Hz to 200~kHz, and the temperature range 100~K $< T <$ 300~K using an HP4284 capacitance bridge. The capacitance was sequentially sampled at 185 frequencies spaced evenly on a logarithmic scale across our bandwidth as the temperature was lowered at a rate of 0.1 K/min, thus guaranteeing a complete frequency sweep over every 0.25~K temperature interval. The capacitances of individual frequencies were then interpolated onto a standard temperature grid with steps of 1K for each frequency, allowing each dielectric spectrum to be analyzed at constant temperature. As a check, the interpolated capacitance values from the multiple-frequency temperature sweep were compared to single-frequency temperature sweeps at several representative frequencies across the bandwidth, and were found to be identical. Warming runs were also performed with no qualitative change in model parameters other than a hysteretic shift in temperature.

\begin{figure}
\includegraphics[angle=0,width=0.45\textwidth]{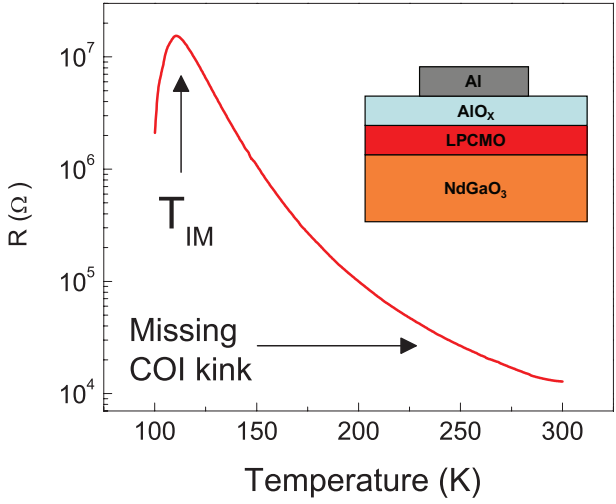}
\caption{Resistance and Sample Geometry. The \textit{a}-\textit{b} in-plane resistance (measured upon cooling) shows a pronounced peak at the insulator-to-metal transition, $T_{IM} \approx 115K$, but lacks a COI associated anomaly seen in bulk manganites in the temperature range 200-250~K\cite{COI_T,TCO1,TCO2}. Inset: Sample geometry, where the LPCMO film is the bottom electrode of a tri-layer capacitor structure.}
\label{fig:RvT}
\end{figure}

\section{\label{analysis}III Analysis}
\subsection{\label{sec:circuit}A. Circuit Model}

\begin{figure}
\includegraphics[angle=0,width=0.5\textwidth]{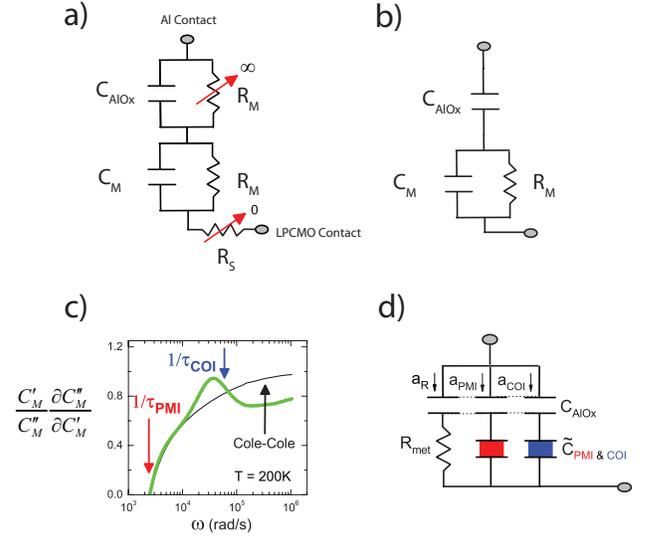}
\caption{ Circuit Models. a) Circuit model composed of a Maxwell-Wagner circuit with a series resistance ($R_s$) component that accounts for the in-plane voltage drop along the film. In select frequency ranges, this in-plane voltage drop can be shown to be negligible compared to the c-axis voltage drops across the capacitors, effectively removing $R_s$ from the circuit\cite{CMC}. Additionally, the AlO$_{x}$ dc resistance is exceptionally large and can be approximated as infinite. b) The resulting circuit model for our measurement bandwidth after approximations and limits are applied. c) The logarithmic parametric-slope,$(C_{M}^{'}/C_{M}^{''})(\partial C_{M}^{''}/\partial C_{M}^{'})$ (green), is shown to differ from the Cole-Cole\cite{Cole-Cole} dielectric response, providing a signature of multiple phases. d) Circuit model incorporating the complex dielectric function of $C_{M}$, which has multiple components. Three parallel components, $C_{PMI}$, $C_{COI}$, and $R_{met}$, are in series to fractional areas of the AlO$_x$ layer.}
\label{circuits}
\end{figure}


Figure \ref{circuits} shows the circuit model used to interpret the dielectric data. The basic treatment of the circuit model was introduced in Ref.~\cite{CMC}, however, we review and expand upon its application here. In contrast to the resistance measurements of Fig.~\ref{fig:RvT} where four contacts are made directly to the LPCMO, the dielectric measurements are two terminal, with one contact to the Al electrode and the other to the edge of the LPCMO film. In this configuration the sample geometry can be represented by a series resistance $R_S$, through which in-plane currents flow from the LPCMO contact to the capacitor structure which comprises a Maxwell-Wagner circuit combination of two series-connected parallel combinations of resistors and capacitors. The voltage drops across the respective capacitances of the LPCMO and AlO$_{x}$ films are in the  the \textit{c}-axis direction, perpendicular to the plane of the substrate. Reference~\cite{CMC} introduced a set of frequency-dependent impedance constraints which guarantee that in our measurement bandwidth the in-plane voltage drop across $R_S$ is negligible compared to the \textit{c}-axis voltage drops, thus effectively removing $R_{S}$ from the circuit. The net result is that the equipotential planes corresponding to the measured voltages are parallel to the film-substrate interface and thus sensitive to the \textit{c}-axis capacitance. This is important because it places the AlO$_{x}$ layer (with approximately infinite resistance) directly in series with the \textit{c}-axis capacitance of the manganite film. 

The resulting circuit is shown in Fig.~\ref{circuits}b, a parallel capacitor and resistor (associated with the manganite capacitance and dc resistance) in series with a capacitor (associated with the AlO$_{x}$ layer). The complex capacitance of this total circuit can then be written as:
\begin{equation}
C^{*} = \frac{C_{AlO_{x}}}{1+(i\omega R_{M}C_{AlO_{x}})/(1+i\omega R_{M}C_{M})}
\label{eq:MWcirc}
\end{equation}
where $C_{AlO_{x}}$ is the capacitance of the AlO$_{x}$ layer, $C_{M}$ is the capacitance of the manganite film, $R_{M}$ is the shorting dc resistance of the manganite film, and $\omega$ = 2$\pi$f: where f is the measurement frequency. At sufficiently high frequency (i.e., when $\omega R_{M}C_{M} >>$1), the capacitance of the circuit may be written:
\begin{equation}
C^{*} = \frac{C_{AlO_{x}} C_{M}}{C_{AlO_{x}} + C_{M}},
\label{eq:hflimit}
\end{equation}
and if $C_{M}$ $<<$ $C_{AlO_{x}}$, as it is in our system, then $C^{*} \approx C_{M}$. We note that $C_{M}$ is a complex capacitance which may display dispersion, but its frequency dependence is independent of $R_{M}$.

The circuit in Fig.~\ref{circuits}b has three dominant time-scales (or frequency ranges). At the longest time-scales, the circuit is dominated by the AlO$_{x}$ layer and $C^{*}$ $\approx$ $C_{AlO_{x}}$ (this can be seen by setting $\omega$ = 0 in Eq. \ref{eq:MWcirc} above). At intermediate time-scales, where 1/$\omega$ $\approx$ $R_{M}$$C_{AlO_{x}}$, the circuit is in a crossover region where C$_{AlO_{x}}$, $C_{M}$, and $R_{M}$ all contribute to the dielectric response. And finally, at time-scales shorter than $R_{M}$$C_{M}$, the capacitance is sensitive only to $C_{M}$, and is independent of the parallel shorting resistor, $R_{M}$, and the AlO$_{x}$ capacitance, $C_{AlO_{x}}$. 

Reference~\cite{CMC} shows a clear crossover between these three time-scales/frequency ranges, and demonstrates that our measurements are made in the high frequency range, where the capacitance is sensitive only to the inherent dielectric properties of the manganite film, $C_{M}$, and not its dc resistance $R_{M}$. Therefore, the time dependence reported is not the result of RC time constants (these occur at lower frequencies), but rather intrinsic relaxation time-scales. Said in another way, there is an essential difference between a \lq\lq lossy\rq\rq capacitor ($C_M$) which is complex and includes the real and imaginary parts of polarization response and a \lq\lq leaky\rq\rq capacitance ($C_M$ in parallel with $R_M$) which provides a shunting path for dc currents. As explained above, the choice of frequency range in our experimental configuration allows us to ignore the contributions of currents flowing through $R_M$ and thus measure the inherent dielectric relaxation of the LPCMO film.

\begin{figure}
\includegraphics[angle=0,width=0.4\textwidth]{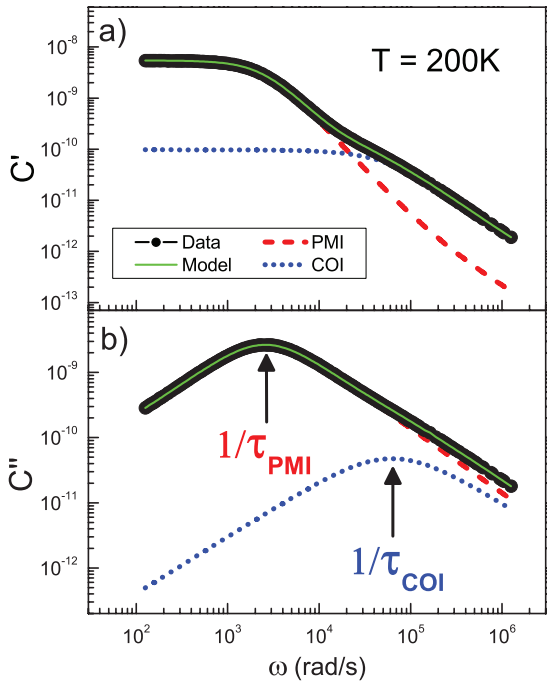}
\caption{Parallel circuit model describes complex capacitance data. a, The measured real capacitance (black) is compared with fits to Eq.~\ref{eq:model} (green) at $T$ = 200~K. b, Measured imaginary capacitance. The individual capacitances of the PMI (red dash) and COI (blue dot) phases are also shown. At low frequencies, the PMI phase dominates both channels, and at high frequencies the COI dominates in the real channel, but not the imaginary channel. This mixing of relaxation components at high frequencies results in the logarithmic parametric-slope behavior seen in Fig.~\ref{circuits}c.}
\label{fig:fits}
\end{figure}

Analyzing the frequency response of the manganite capacitance ($C_{M}$) reveals a high-frequency anomaly as shown in Fig.~\ref{circuits}c, which compares our capacitance data to the ubiquitous\cite{universal} Cole-Cole dielectric response\cite{Cole-Cole} of standard dielectric theory. Plotting the logarithmic parametric slope vs. frequency, $(\partial (\textrm{ln}~C^{\prime \prime}_{M})/\partial \omega)/(\partial (\textrm{ln}~C^{\prime}_{M})/\partial \omega)$, (where $C^{\prime}_{M}$ and $C^{\prime \prime}_{M}$ are the real and imaginary capacitances with $C^{\prime \prime}$ = 1/$\omega R_{P}$ directly related to the parallel resistance $R_{P}$ reported by the capacitance bridge) the Cole-Cole response increases monotonically while our data display a high-frequency non-monotonic anomaly. The zero crossing in Fig.~\ref{circuits}c represents the loss peak (see also Fig.~\ref{fig:fits}) of the PMI phase, and considering the phase coexistence found in bulk, we postulate that the high frequency anomaly in $C_{M}$ is due to a higher frequency dielectric relaxation of the COI phase. 

To test our hypothesis that the COI phase is responsible for the high frequency anomaly, we modify the circuit model of Ref.~\cite{CMC} to account for multiple phases.
Figure~\ref{circuits}d shows the modified circuit in which $C_{M}$ encompasses phase coexistence. It is composed of three parallel components all in series to fractional areas of the AlO$_x$ dielectric layer: $a_{COI}$, $a_{PMI}$, and $a_{R}$ (the fractional area of the FMM phase, which acts as a resistive short at low temperatures), with the constraint $\sum{a_{i}} = 1$. Placing the capacitances of each phase in parallel requires that the domains of each phase span the film thickness, thus  obviating a series configuration. Our film thickness of 30~nm, however, likely satisfies this requirement, as the phase domains of manganites in multiple phase separation states have been shown to be on the order of microns\cite{MFM,MPS}. Above $T_{IM}$, $a_{R} \approx 0$, and the circuit is dominated by $C_{PMI}$ and $C_{COI}$ in our frequency range, so that the total dielectric response may be approximated by the superimposition of two Cole-Cole dielectric functions, 
\begin{equation}
\tilde{\epsilon}(\omega)\  =\  \epsilon_{\infty} +\frac{A_{PMI}}{1 + (i \omega \tau_{PMI})^{1-\alpha}} + \frac{A_{COI}}{1 + (i \omega \tau_{COI})^{1-\beta}},
\label{eq:model}
\end{equation}
where the amplitudes $A_i$ are the product of the fractional area $a_i$ and dielectric constant $\epsilon_i$ of each phase ($A_{i} = a_{i}\epsilon_{i}$), $\epsilon_{\infty}$ includes the infinite frequency response of both dielectric phases, $\tau_{COI}$ and $\tau_{PMI}$ are the respective relaxation time-scales, and $\alpha$ and $\beta$ are the respective dielectric broadenings.

We quantitatively model $C_{M}$ with Eq.~\ref{eq:model} using the fixed temperature dielectric spectrums, varying $\omega = 2\pi f$ over 185 frequencies over the bandwidth 20~Hz to 200~kHz (see Sec. IIB). The bandwidth was chosen such that at the lowest frequencies Eq. \ref{eq:hflimit} is still valid, and that at the highest frequencies the transverse a-b plane series resistance $R_{S}$ can still be ignored. The fits are produced by simultaneously minimizing the difference between the measured complex capacitance and both the real and imaginary parts of Eq.~(2), over 370 independent data points in total. Figure~\ref{fig:fits} shows a typical fit, where the average relative error is less than $10^{-3}$. In the low-frequency limit, $\epsilon(0) \approx A_{PMI} + A_{COI}$, allowing a fitting variable to be eliminated by reparameterizing the dielectric amplitudes in terms of their ratio, $r_{amp} =A_{COI}/A_{PMI}$, and the measured $\epsilon(0)$, i.e., $A_{PMI} = \epsilon(0)/(1+r_{amp})$,and $A_{COI} = r_{amp}~\epsilon(0)/(1+r_{amp})$. As $\tau_{PMI}$ is determined from the low frequency loss peak (see Fig.~\ref{circuits}c and Fig.~\ref{fig:fits}b), five free variables are determined from the 370
independent data points: $\epsilon_{\infty}$, $r_{amp}$, $\alpha$, $\beta$ and $\tau_{COI}$. We fit our complex capacitance data to this five-parameter model (Eq.~\ref{eq:model}) at fixed temperatures in 1~K steps between 100~K and 300~K. Analyzing the temperature dependence of the model parameters permits the identification of the phases, and provides a detailed spatial and temporal characterization of their coexistence/competition.

\subsection{B. Temperature Dependence of Model Parameters}

\begin{figure}
\includegraphics[angle=0,width=0.40\textwidth]{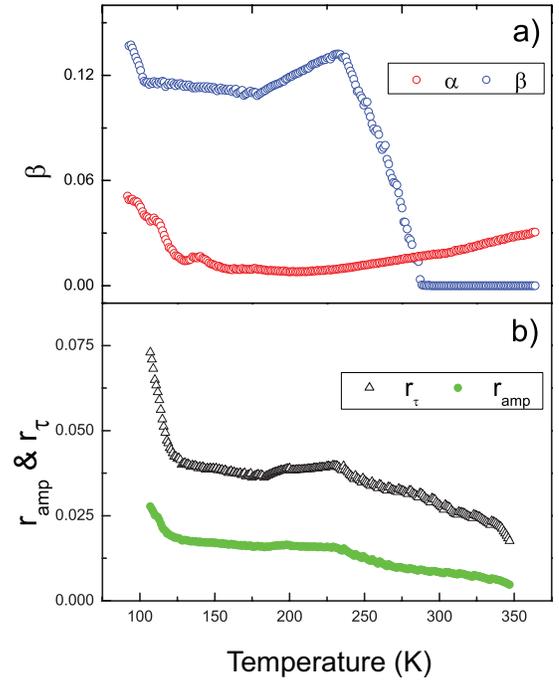}
\caption{Temperature dependence of selected model parameters identifies phases and verifies detailed balance. a) The temperature dependence of $\beta$ is shown to match the temperature dependence of the correlations of COI phase, where COI nanoclusters are reported in a related material to appear near 280~K with the phase fully developed below 240~K\cite{COI_T}, identifying the high-frequency relaxation as the COI phase. The broadening of the PMI phase, $\alpha$, is shown to be featureless in this temperature range. b) The temperature dependence of the ratios of the dielectric amplitudes, $r_{amp} =$ $a_{COI}/a_{PMI}(\epsilon_{COI}/\epsilon_{PMI})$, and time-scales, $r_{\tau} = \tau_{COI}/\tau_{PMI} = a_{COI}/a_{PMI}$, are compared. As discussed in the text, the strong correlation between the ratios confirms that the phase competition is governed by detailed balance.}
\label{fig:paramT}
\end{figure}

Dielectric broadening provides a measure of the correlations among relaxors, and it is known that the COI phase is a highly correlated and ordered phase. Thus, it is expected that the broadening of the COI phase should increase as the phase forms. The temperature dependence of $\beta$ displays these charge-ordering features (see Fig.~\ref{fig:paramT}a), while $\alpha$ is featureless, thereby identifying the high-frequency response as the COI dielectric phase. As seen in Fig.~\ref{fig:fits}, the COI phase dominates the high-frequency response of the real capacitance, but only contributes slightly to the imaginary capacitance.  Thus, the high frequency features of the logarithmic parametric slope (seen in Fig. \ref{circuits}c are the result of a mixture of the real component of the COI phase and the imaginary component of the PMI phase, and are a signature of phase separation. Figure \ref{fig:paramT} also shows the temperature dependnce of the ratio of dielectric amplitudes (discussed in detail below).

\begin{figure}
\includegraphics[angle=0,width=0.40\textwidth]{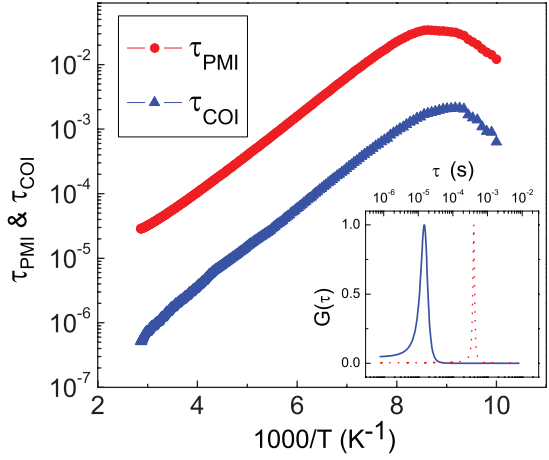}
\caption{Region of identical activation energies provides clue for detailed balance. Arrhenius plots are shown for $\tau_{COI}$ and $\tau_{PMI}$, with $E_{A}\ \approx\ 118$~meV in the linear region for both dielectric phases. The nearly identical $E_{A}$'s suggest the phases share a single energy barrier (depicted in the middle panel of Fig.~\ref{figDB}b). Inset: The distributions of hopping-rate time-scales are shown to be narrow for both phases, suggesting temporally coherent hopping (PMI dotted red, and COI solid blue).}
\label{fig:arrhenius}
\end{figure}

Figure \ref{fig:arrhenius} shows Arrhenius plots of $\tau_{PMI}$ and $\tau_{COI}$ over the temperature range 100~K~$ < T < $~350~K. Surprisingly, over the linear regions, the activation energies of each phase are nearly equal, with $E_{A}(PMI) = 117.9 \pm 0.2$~meV and $E_{A}(COI) = 118.6 \pm 0.3$~meV. These values are consistent with small polarons, the known conduction and polarization mechanism in manganites\cite{Mang_Review,mdcon}. 

\subsection{C. Detailed Balance}

The strikingly similar activation energies of the two phases suggests the relaxations are coupled, possibly sharing a common energy barrier. Crossing this energy barrier would result in the two phases converting into each other. In our system, however, each dielectric also polarizes independently without converting into the other phase.  Therefore, the phases must be connected through a common excited state from which relaxations can occur to either phase. This common state is consistent with adiabatic polaron hopping in manganites\cite{Mang_Review}, where the lattice relaxes slowly in response to fast electronic hopping (a polaron is a quasi-particle that includes an electron and the lattice distortion caused by its presence).

\begin{figure}
\includegraphics[angle=0,width=0.7\textwidth]{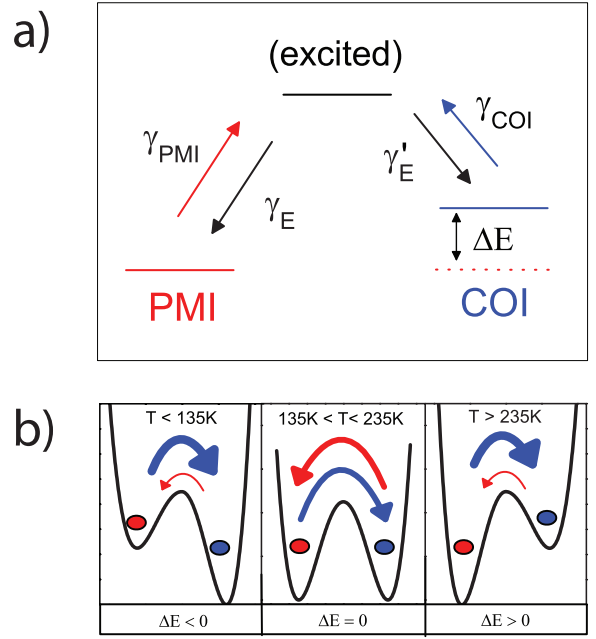}
\caption{ Detailed Balance Model. a) An energy level schematic of a three state system describing the polaron hopping process is presented. Polarons of both dielectric phases absorb thermal fluctuations and hop to an excited state at their characteristic rates, $\gamma_{PMI}$ and $\gamma_{COI}$.  The excited state is equivalent for both phases, corresponding to an electron surrounded by an undistorted lattice. The lattice can then relax into distortion states that correspond to polarons of either dielectric phase. b) Schematic depiction of the temperature evolution of detailed balance and the energy barriers separating the PMI and COI dielectric phases.}
\label{figDB}
\end{figure}

We model this process in our samples by the three state system shown in Fig.~\ref{figDB}. The electrons of the polarons of both dielectric phases absorb thermal fluctuations that activate them over their hopping barriers to an equivalent ``excited" state: a relocated electron surrounded by a lattice site that has yet to relax. The new lattice site has some initial distortion (either PMI or COI), but as it accommodates the new electron it can transform/relax into distortions that correspond to either dielectric phase. The electronic hopping happens at characteristic rates which we measure directly from loss peak positions in the complex capacitance ($\gamma_{P} = 1/\tau_{PMI}$, and $\gamma_{C} = 1/\tau_{COI}$). The lattice site relaxation, however, occurs at unknown rates, $\gamma_{E}$ and $\gamma_{E}^{'}$, for the PMI and COI phases respectively. This process effectively results in two channels, one in which polarization is manifested independently in each phase by polarons relocating without altering their distortions, and one in which polarons relocate as well as transforming their distortion state.

Since the equilibrium populations of each phase are constant in time, the rate equations for the three state model in Fig.~\ref{figDB} are given by,
\begin{equation}
\frac{\partial}{\partial t}
\begin{pmatrix}
n_{P}\\
n_{E}\\
n_{C}
\end{pmatrix}
=\begin{pmatrix}
-\gamma_{P} & \gamma_{E} & 0 \\
\gamma_{P} & -(\gamma_{E}+\gamma_{E}^{'}) & \gamma_{C}\\
0 & \gamma_{E}^{'} & -\gamma_{C}\\
\end{pmatrix}
\begin{pmatrix}
n_{P}\\
n_{E}\\
n_{C}
\end{pmatrix}
\end{equation}
where $n_{P}$, $n_{C}$, and $n_{E}$ are the populations of the PMI, COI, and excited state respectively. Solving this system of equations at equilibrium results in a detailed balance equation of the form,
\begin{equation}
n_{C}(\gamma_{C}\gamma_{E})  = n_{P}(\gamma_{P}\gamma_{E}^{'})~,
\end{equation}
where ($\gamma_{C}\gamma_{E}$) and ($\gamma_{P}\gamma_{E}^{'}$) are the effective transition probabilities of each phase. Although the populations of each phase are time-independent at equilibrium, they still have inherent temperature and energy dependences governed by Boltzmann statistics. The populations of each phase may be written in terms of their ground-state population and an exponential factor, 
\begin{equation}
\begin{array}{clr}
n_{P} = n_{P}^{0} e^{-E_{PMI}/kT} \\
n_{C} = n_{C}^{0} e^{-E_{COI}/kT}
\end{array}
\label{eq:pop_ratio}
\end{equation}
where $E_{PMI}$ and $E_{COI}$ are the configuration energies of each phase. We stress here the distinction of $E_{A}(i)$ and $E_{i}$ with ($i = PMI, COI$). $E_{A}(i)$ is the energy barrier to hopping, and is thus the energy difference between the current polaron state and the excited energy state: $E_{A}(i) = E_{excited} - E_{i}$ with ($i = PMI, COI$). The detailed balance equation may then be rewritten as, 
\begin{equation}
(n_{C}^{0}/n_{P}^{0})e^{-\Delta E/kT} = (\tau_{COI}/\tau_{PMI})(\gamma_{E}^{'}/\gamma_{E})~,
\label{eq:detailed_bal}
\end{equation}
where $\Delta E = E_{COI} - E_{PMI}$ is the difference in configuration energy between phases.

By making the physically reasonable ansatz that the ratio of populations is equal to the ratio of fractional areas (i.e., volumes for constant thickness),
\begin{equation}
(n_{C}^{0}/n_{P}^{0})e^{-\Delta E/k_BT} = a_{COI}/a_{PMI}~~,
\label{eq:assumption}
\end{equation} 
our circuit model provides a direct test of the detailed balance constraint of Eq.~\ref{eq:detailed_bal}. Figure~\ref{fig:paramT}b shows the temperature dependence of the \textit{independently determined} ratios, $r_{\tau} =$$\tau_{COI}/\tau_{PMI}$ and $r_{amp} =$ $(a_{COI}/a_{PMI})(\epsilon_{COI}/\epsilon_{PMI})$. The two ratios follow a similar trend with a ratio of ratios, $r_{\tau}/r_{amp}=$ $(\tau_{COI}/\tau_{PMI})(a_{PMI}/a_{COI})(\epsilon_{PMI}/\epsilon_{COI})\approx 2.4$ over the displayed temperature range. 

Combining Eqs.~\ref{eq:detailed_bal} and ~\ref{eq:assumption} leads to the simplified expression,
$r_{\tau}/r_{amp} =(\epsilon_{PMI}/\epsilon_{COI})(\gamma_{E}/\gamma_{E}^{'})$ which is confirmed to be a constant with additional measurements for different thickness films of $\epsilon_{COI}$ and $\epsilon_{PMI}$ that are found to be in agreement with bulk values, providing the result $\gamma_{E}^{'}/\gamma_{E} \approx 1$ (see below). The similarities in the temperature dependence of $r_{\tau}$ and $r_{amp}$ in Fig.~3b thus confirm the constraints imposed by detailed balance. 

\begin{figure}
\includegraphics[angle=0,width=0.4\textwidth]{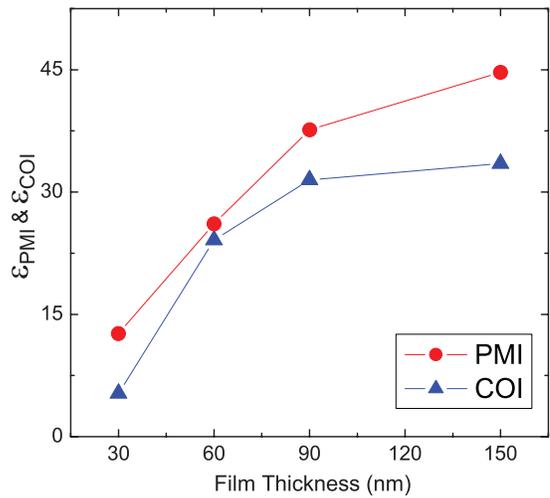}
\caption{Thickness Dependence of Dielectric Constants. The dielectric constants, calculated from the dielectric amplitudes and areas of each phase, $\epsilon_{i} = A_{i}/a_{i}$,  are shown to saturate near their bulk values once substrate strain is relaxed. The agreement with bulk data validates the areas calculated assuming equal lattice relaxation rates, $\gamma_{E} \approx \gamma_{E}^{'}$.}
\label{fig:thickness}
\end{figure}

Assuming the lattice relaxation rates are equal ($\gamma_{E}^{\prime}/\gamma_{E} \approx 1$, this assumption is validated below), then according to Eqs. \ref{eq:detailed_bal} and \ref{eq:assumption} the ratio of fractional areas is known. Knowledge of the ratio of fractional areas $a_{COI}/a_{PMI}$ and the ratio of dielectric amplitudes $A_{COI}/A_{PMI}$ together with the respective constraining normalizations, $a_{COI} + a_{PMI} = 1$ and $A_{COI} + A_{PMI} = \epsilon(0)$, allow an experimental determination of the respective dielectric constants $\epsilon_{COI} = A_{COI}/a_{COI}$ and $\epsilon_{PMI} = A_{PMI}/a_{PMI}$. Figure~\ref{fig:thickness} shows the dielectric constants determined in this manner for four films with thickness ranging from 30~nm to 150~nm. The dielectric constants of each phase increase and saturate near their known bulk values\cite{orig,mdcon,epmi} as the substrate strain relaxes. In manganites grown on NGO, the film is relaxed at a thickness of $d \approx 100$ nm\cite{thick,CMC} in agreement with the saturation of the data in the figure. The agreement of the data with these expectations tends to validate our assumption that $\gamma_{E}^{'}/\gamma_{E} \approx 1$.

The combination of Eq.~\ref{eq:detailed_bal} and Eq.~\ref{eq:assumption} together with the result that $\gamma_{E}^{'}/\gamma_{E} \approx 1$  leads to the relation $a_{COI}\gamma_{COI} \approx a_{PMI}\gamma_{PMI}$ which with the normalization, $a_{COI} + a_{PMI} = 1$, gives the particularly simple relations,
\begin{equation}
\begin{array}{clr}
a_{COI} \approx \gamma_{P}/(\gamma_{C} + \gamma_{P}), \\ 
a_{PMI} \approx \gamma_{C}/(\gamma_{C} + \gamma_{P}),
\label{eqs:aCOI+aPMI}
\end{array}
\end{equation}
for the fractional areas occupied by each phase. 

The consequences of our three state model as expressed in the above equations and illustrated in the schematic panels of Fig.~\ref{figDB}b give a good description of the temperature evolution of the cooling data shown in Fig.~\ref{fig:paramT}b. As the COI phase stabilizes with decreasing temperature ($T > 235$~K), its hopping rate decreases (polarons remain in the state longer because of a deeper potential well), increasing its population proportional to $e^{-\Delta E/kT}$, since $\Delta E \rightarrow 0$ as $E_{COI}$ decreases. This populations change is depicted by the colored arrows, the size of which represent the magnitude of the respective transition rates. Then at intermediate temperatures (135~K$<T<$235~K) the two phases are at equal energies ($\Delta E = 0$), making their relative populations temperature independent (equal size arrows and transition rates). Finally, at low temperatures the PMI phase destabilizes and $\Delta E$ decreases further and becomes negative as $E_{PMI}$ increases (as shown in Fig.~\ref{figDB}b), resulting in an increase of the COI population proportional to $e^{-\Delta E/kT}$.

\subsection{D. Charge Density Waves}

The original prediction of `electronically soft matter' describes the COI phase in terms of charge-density waves (CDWs)\cite{ESP}. Since evidence for CDWs in manganites has been verified by multiple experimental techniques \cite{CDW_sliding,CDW_opt,CDW_noise,CDW_Fisher}, it is therefore appropriate to discuss our data in this context. With respect to electrical measurements of resistivity and noise, the preponderance of evidence\cite{CDW_sliding,CDW_noise} points to delocalized CDWs that slide in response to applied electric fields (i.e., sliding CDWs). Although the work presented here is not able to resolve opinion as to whether CDWs in the COI phase are localized\cite{CDW_Fisher} or delocalized\cite{CDW_sliding,CDW_noise}, we are nevertheless able to utilize the widely accepted CDW picture to provide considerable insight into the dynamics between coexisting phases, one of which (the COI phase) contains charge disproportionation in the form of CDWs. 

We first note that the absence of the COI transition in the resistance vs. temperature curve (despite the observed phase competition) is analogous to similar behavior in CDW systems doped with large impurity densities\cite{DCDW1,DCDW2}. The lack of a feature does not indicate the phases absence, rather it is the smearing of the COI transition by the inherent disorder and strain of thin films\cite{CDW_sliding}. Furthermore, characterizing the temperature dependence of the PMI/COI competition reveals a highly correlated collective transport mode of the COI phase domains, similar to the `coherent creep' preceding `sliding' in CDW systems\cite{Coherent_Creep}.


The constraints of the parallel model
(Fig.~\ref{circuits}d and Eq.~\ref{eq:model})
require that the hopping mechanism is 
correlated over sufficiently long length scales that regions equal to at least the film thickness hop together collectively, so that as the phases convert each phase boundary progresses simultaneously in a `creep' like manner. `Creep' is typically a random phenomenon, however, transforming our dielectric broadening to a distribution of time-scales\cite{DRT} (shown in the inset of Fig.~\ref{fig:arrhenius}) we find a narrow distribution of hopping rates suggesting an ordered process similar to the `temporally coherent creep' found in the CDW system NbSe$_{3}$\cite{Coherent_Creep}. The exact nature of the order is ambiguous, with two likely scenarios. The first possibility is the coherent propagation of phase domains, where as the phase boundary `creeps' forward the regions behind synchronously hop, guaranteeing the continuity of the phase. The second scenario is a `breathing' mode in which the area of different phase domains cooperatively increase and decrease at a characteristic frequency (with total area conserved). Both scenarios demonstrate the collective and delocalized nature of the COI phase in which its entire charge distribution moves collectively and coherently in dynamic competition with the PMI dielectric phase.

\section{IV. Conclusions}

In summary, we have presented a dielectric characterization of the competition between the COI and PMI dielectric phases of (La$_{1-y}$Pr$_{y}$)$_{0.67}$Ca$_{0.33}$MnO$_{3}$, identifying signatures of phase separation and providing temperature dependent time-scales, dielectric broadenings, and population fractions of each phase. More importantly, we demonstrate that the constraints imposed by detailed balance describe an `electronically soft' coexistence and competition between dielectric phases, highlighted by continuous conversions between phases on large length and time scales as well as a collective and delocalized nature of the charge-density distribution of the COI phase. Our findings provide important context concerning the fundamental mechanisms driving phase separation, and strongly support the concept of an \lq\lq electronically soft\rq\rq separation of delocalized competing thermodynamic phases\cite{ESP}. Furthermore, we extend this concept to high temperature fluctuating phases which need not be ordered.

\section{Acknowledgments}
The authors thank Tara Dhakal, Guneeta Singh-Bhalla, Chris Stanton and Sefaatin Tongay for assistance with sample preparation as well as fruitful discussions. This research was supported by the U.S. National Science Foundation under Grant No. DMR-1005301 (AFH) and No. DMR 0804452 (AB).

\end{document}